\documentstyle[prl,preprint,aps]{revtex}
\begin{document}
\draft
\title{Composite Fermions in Quantum Dots}
\author{J.K. Jain and T. Kawamura}
\address{Department of Physics, State University of New York
at Stony Brook, Stony Brook, New York 11794-3800}
\date{\today}
\maketitle
\begin{abstract}

We demonstrate the formation of composite fermions in
two-dimensional quantum dots under high magnetic fields. The
composite fermion interpretation provides a simple way to understand
several qualitative and quantitative
features of the numerical results obtained earlier in exact
diagonalization studies. In particular, the
ground states are recognized as compactly
filled quasi-Landau levels of composite fermions.

\end{abstract}

%\narrowline

Composite fermions \cite {Jain} provide simple insight into otherwise
intriguing
phenomena observed in two-dimensional electron systems (2DES) in high
magnetic fields ($B$). The fractional quantum Hall effect (FQHE) \cite
{Tsui} is understood as the integer quantum Hall effect of composite
fermions \cite {Jain}, and the experimental properties of the
compressible state at
the half-filled Landau level (LL) are explained in terms of a Fermi
sea of composite fermions \cite {HLR}.
In earlier studies, the composite fermion (CF)
theory has been tested numerically in geometries without edges, to
simulate the bulk of a large 2DES, and found to work extremely well
\cite {DWJ}. This work investigates its  applicability to a two-dimensional
quantum dot in high $B$, containing a few electrons in the presence of
a confining potential. The charge density is expected to be
non-uniform, the (local) filling factor is not necessarily a
useful concept, and it is {\em a priori} not obvious that composite
fermions should be formed in this system.

Two-dimensional semiconductor quantum dots containing a few
electrons have recently been realized experimentally, and studied
by tunneling \cite {Expt1}, capacitance \cite {Expt2}, and optical
\cite {Expt3} spectroscopy. Since these measurements probe the energy
levels of the many-body system of interacting electrons,
there have been several theoretical studies calculating
the eigenspectrum as a function of the confining potential, the number of
electrons ($N$), and the magnetic field ($B$). It  has been found that
the system exhibits a rather rich and complex energy spectrum \cite
{Tapash,Girvin,Yang,Others,Stone}. However, no systematic
way of understanding these results exists. This paper shows
that the CF theory provides a simple and coherent explanation of
a number of qualitative and quantitative features of the numerical
results.

Following the usual practice, we consider in our numerical studies
a model in which electrons, restricted to two
dimensions, are confined by a parabolic potential of the form
$\frac{1}{2}m^* \omega_{0}^2 r^2$, where $m^*$ is the effective
mass of the electron. The single electron problem is essentially
that of a harmonic oscillator, and can be solved exactly
\cite {Fock} for arbitrary $B$.
This paper considers the limit of $B\rightarrow \infty$, so
that the single particle states are labeled by a
LL-index $n=0,1,...$ and an angular momentum index $l=-n,-n+1,...$.
The electrons are fully polarized, and occupy only the lowest
LL (LLL). The Coulomb interaction conserves the total angular momentum
$L=\sum_{i}l_{i}$, which will be used to label the many-body
eigenstates.  In the limit $B\rightarrow\infty$ the
eigenenergies of the many body system conveniently separate into an
interaction part, $V(L)$, and a confinement part, $E_{c}(L)$:
$E(L)=E_{c}(L)+V(L)$. The confinement energy has the simple form
$E_{c}(L)=(\hbar/2)[\Omega-\omega_{c}] L$,
relative to the LLL, where $\omega_{c}=eB/m^*c$ is the cyclotron
frequency, and $\Omega^2=\omega_{c}^2+4\omega_{0}^2$.
The interaction energy $V(L)$ is the same as that of electrons
without confinement, but at a magnetic field ${\cal B}=\Omega m^*c/e$.
Therefore, it is sufficient to consider interacting electrons in
the absence of any confining potential to determine $V(L)$.
$V(L)$ will be expressed in units of $e^2/\epsilon a$
where $a=\sqrt{\hbar c/{e \cal B}}$ and $\epsilon$ is the
dielectric constant of the background material.
Figs. 1-3  shows $V(L)$ (for the lowest-energy state with a given $L$)
as a function of $L$ for $N=3$-6.  (The reason for the seemingly
peculiar choice of the horizontal scale will become clear
later.) Such curves have been obtained in a number of earlier
calculations, but to our knowledge, no general
intuitive way of understanding the shape of the curves, the
positions of the cusps, or the structure of the states has been
provided. We
now investigate this problem in the context of the CF theory.

A composite fermion is obtained when an electron captures an even
number ($2m$) of  vortices of the wave function.
Microscopically, the formation of composite fermions implies that the
(unnormalized) low-energy wave functions of interacting
electrons with total angular momentum $L$ have the form:
\begin{equation}
\Psi_{L}={\cal P}\prod_{j<k}(z_{j}-z_{k})^{2m}\Phi_{L^*}\;\;,
\label{cfwf}
\end{equation}
\begin{equation}
L=L^*+mN(N-1)=L^*+2mM\;\;,
\label{cfL}
\end{equation}
where $\Phi_{L^*}$ is the wave function of {\em non-interacting}
electrons with total angular momentum $L^*$, $z_{j}=x_{j}-iy_{j}$
denotes the position of the $j$th electron, and
${\cal P}$ projects the wave function on to the LLL of
electrons, as appropriate for $B\rightarrow\infty$.
The Jastrow factor binds $2m$ vortices to each electron of
$\Phi_{L^*}$ to convert it into a composite fermion.
We restrict $L^*$ to the range $-M\leq
L^* \leq M$, where $M=N(N-1)/2$; the CF theory relates interacting
electrons at arbitrary $L$ to non-interacting electrons
in this range with a suitable choice of $m$.
At $L^*=-M$, the lowest-kinetic-energy state contains one electron in the
lowest angular momentum state of each of the lowest $N$  LL's.
The kinetic energy of this state is given by $K(-M)=M\hbar\omega_{c}$
(relative to the LLL).  The lowest-energy state at $L^*=M$ contains
all electrons in the LLL, with kinetic energy equal to zero.
[The CF states at the corresponding $L=(2m+1)M$ are identical to
the Laughlin
states \cite {Laughlin}.] The $K(L^*)$ in the entire $L^*$ range
is plotted in Figs. 1-3 (the dashed line).  (The cyclotron energy
has been replaced by a quasi-cyclotron
energy, the value of which will be obtained later. At the moment, we
only consider the shape of the curve.)
Clearly, the shape of $V(L)$, shifted by  $2mM$,
is remarkably similar to that of $K(L^*)$ \cite {Rejaei}.
In particular, the positions of all upward and downward cusps are
predicted correctly.  Note that we actually need the $L^*$ spectrum
only from 0 to $M$; the other half (0 to $-M$) can be obtained
from a reflection symmetry of the problem:
the degeneracies at $L^*$ and $-L^*$ are same; the wave functions
are related by complex conjugation; the horizonatal
lines of one map into tilted lines of the other and vice versa (Figs.
1-3); and a cusp at $L^*$ implies a cusp at $-L^*$.

An important theoretical goal in the quantum dot physics is to
identify the (global) ground state as a function of $N$, $B$, and
the strength of the confining potential, investigated in detail in Ref.
\cite {Yang}. It occurs at the minimum
of $E_{c}(L)+ V(L)$, and {\em is in general one of those states where there
is a downward cusp}. For a fixed $B$, as the strength of the confining
potential is increased, the area occupied by the electrons will
decrease, and the ground state will progressively shift to
downward-cusp  states with smaller and smaller $L$.
For example, for $N=6$, ground states were found
in \cite {Yang} to occur at $L=$ 15, 21, 25, 30, 35, 39, and 45
(at large Zeeman energy) as the strength of the confinement
potential was varied.  There are downward cusps at all corresponding $L^*$.
Similarly, for $N=5$, ground states occur at $L=$ 10, 15, 20, 25, and
30, which again correspond to downward cusps \cite {F5}.

All possible ground states states possess a simple
microscopic structure in terms of composite fermions.  To see this,
let us again start by considering non-interacting electrons in the
range
$-M\leq L^* \leq M$ and denote the number of electrons in
the $n$th LL by $N_{n}$ ($\sum_{n}N_{n}=N$).  We then define ``compact
states" as those states in which  the $N_{n}$ electrons fill the
innermost ($l=-n, -n+1, ..., -n+N_{n}-1$) single-particle
states of the $n$th LL.  We denote the compact states by
$[N_{0},N_{1},... N_{s}]$, where $s$ is the index of the highest
occupied LL.  It is easy to see that {\em all possible ground
states are compact}.
(The inverse is clearly not true.) For other states, the
total angular momentum can be reduced without increasing the kinetic
energy by simply shifting the angular momentum of the
electrons within the same LL; i.e., there can be no downward cusp.
Thus, in determining the ground states of
non-interacting electrons, it is sufficient to consider only the
compact states. Figs. 1  and 2 display the compact state associated
with each cusp; the two extreme states are $[N]$ and $[1,1,...,1]$.

The ground states of {\em interacting} electrons are then compact states
of {\em composite fermions}, denoted by $[N_{0},N_{1},...N_{s}]_{CF}$.
Their wave functions are obtained according to Eq.\ (\ref{cfwf}),
i.e.,  by  multiplying the wave functions of compact electron states
by
the Jastrow factor, and then projecting on to the LLL. Since the wave
function of a compact electron state is a single Slater determinant,
the wave functions of compact composite fermion states contain
no free parameters. Several compact states were studied in Ref. \cite
{Dev} in a different context. Some of the ground states of the present
study were included there. In particular, the five electron states at
$L=30$ ($[5,0]_{CF}$), $L=25$ ($[4,1]_{CF}$), and $L=22$
($[3,2]_{CF}$) were found to have overlaps of 0.988, 0.993, and 0.998,
respectively, with the corresponding exact ground state. Similarly,
for $N=6$, the composite fermion states at $L=45$ ($[6,0]_{CF}$),
$L=39$ ($[5,1]_{CF}$), and $L=35$  ($[4,2]_{CF}$) had overlaps
of 0.986, 0.994, and 0.991, respectively, with the corresponding
true electron states. Other compact composite fermion
states should provide equally good representations of the exact ground
states.

The above discussion shows that
the CF theory immdiately identifies the possible ground states
out of a large number of available states,
and their microscopic structure.  Now we come to the relative
strengths of various cusps. For non-interacting electrons in
the range $-M\leq L^* \leq M$, the
cusp size, defined by $K(L^*+1)+K(L^*-1)-2K(L^*)$, is the same for all
cusps, equal to the cyclotron energy.  This suggests \cite {HLR}
that all cusps of a given type of composite fermions might
be of (roughly) equal size, with the size of the cusp at $L$ defined
as $E(L+1)+E(L-1)-2E(L)=V(L+1)+V(L-1)-2V(L)$.
We plot in Fig.4 the sizes of all cusps of Figs. 1-3, and also of
those cusps of seven and eight electron systems which we can obtain
numerically. They are indeed comparable for all cusps in a given
$m$ region (for fixed $N$), and can be interpreted as the
quasi-cyclotron energy of composite fermions.
(Some cusps are anomalously weak, e.g. the cusps at $L=12$ and 24 for
$N=4$. These, however, often do not become ground states.)
The cusps of the $m=2$  composite fermions are smaller roughly by
a factor of two to three.  An abrupt change in the cusp size in experiments
might thus be indicative of a transition from one kind of composite
fermions to another.

These results demonstrate the validity of the CF picture,
in which the LL's of electrons in $\Phi$
map into {\em quasi-}LL's of composite fermions,
and the cyclotron energy ($\hbar\omega_{c}$) of electrons in $\Phi$
translates into a quasi-cyclotron energy of composite fermions.
The quasi-cyclotron energy of the $m=1$ composite fermions can
be determined easily by evaluating the size of the
{\em upward} cusp at $L=M+1$, which we denote by
$\hbar\omega^{CF}$. This requires a diagonalization of
only a $2\times 2$ matrix, since the
Hilbert space size at $L=M,\;M+1,$ and $M+2$ is 1, 1, and 2,
respectively. (The calculation is more complicated for the $m=2$
composite fermions.) The cusp sizes in Fig.4 are close to
$\hbar\omega^{CF}$ thus determined \cite {cusp}.
Once $\hbar\omega^{CF}$ is determined,
it is also possible to predict, with some accuracy (which is rather
good close to $\nu=1$), which of the
possible ground states is the actual
ground state for a given set of parameters, and a phase diagram
can be produced.  We will not do that here since it has already been done
for small systems.  Our objective here is to demonstrate that the
CF theory is valid; a CF phase diagram for bigger systems (for which
exact diagonalization is not possible) can be constructed easily
if needed.

We have made above the (lowest-order) approximation
of considering {\em strictly non-interacting} composite fermions,
i.e., we assumed that the Coulomb interaction is {\em completely}
exhausted in the formation of composite fermions.
The residual interaction between composite fermions needs to
be considered for a more detailed understanding. As a result of this
interaction, the composite fermion states, which are degenerate in the
above analysis, will form a band, with the band-width related to the
strength of the residual interaction. This is why the the dashed straight
lines map on to somewhat curved solid lines in Figs. 1-3.
Also, when the residual interaction is comparable to the
quasi-cyclotron energy, which becomes more likely for large $m$,
the composite-fermion description will become less trustworthy, and
some of the weaker cusps will be washed out.

This study resolves several previously noted but unexplained features.
(i) Earlier, there was no general understanding of the
positions or the origin of the cusps. While it was felt that the
physics underlying these cusps was similar to that of the fractional
quantum Hall effect, no explicit connection had been made (with
the exception of Laughlin states \cite {Laughlin}). The present study
shows that the
underlying physics of the cusps here is indeed the same; they are also
a manifestation of the existence of composite fermions.
(ii) It had been noted earlier \cite {Tapash,Girvin} that a subset
of downward cusps appears at an interval of $\Delta L=N$. They
appear, for example, at $L=$ 3, 6, 9 for $N=3$ and $L=$ 6, 10, 14,
18 for $N=4$. These are explained as corresponding to the compact states
$[N]_{CF}$, $[N-1,1]_{CF}$, $[N-2,1,1]_{CF}$, ...,  $[1,1,...1]_{CF}$.
(iii) In Ref. \cite {Dev}, it was found that while some compact states
provide an accurate description of the actual ground state, some do
not. For example, $[3,3]_{CF}$ and  $[4,4]_{CF}$ have rather poor overlaps
with the
actual ground states at the corresponding angular momenta. This was
not understood at the time, but the reason is now clear: these states
are {\em not} non-degenerate; $[3,3]$ is degenerate with
$[4,1,1]$, and $[4,4]$ is not even a (possible) ground state.
(iv) The Laughlin states, $[N]_{CF}$, were first studied in the disk
geometry \cite{Laughlin}. This geometry, however, has not proven useful for
other FQHE states, despite several attempts (see, e.g.,
\cite {Girvin,Dev}). The reason is that, with the exception of
$[N]_{CF}$, no filling factors may be associated with
the cusps in the thermodynamic limit; they simply occur as
composite fermions are promoted one
by one to higher quasi-LL's. To understand this better,
consider 2/5, which one would like to identify with $[M,M]_{CF}$.
However, there is no cusp at $[M,M]$ (for $M>3$), since states with
the same (kinetic) energy can be constructed at smaller angular momenta
(e.g. $[M+1,M-2,1]$). This is possible since an
arbitrary number of electrons can be put in the LLL.
If the size of the system were restricted, so that
each LL had only a {\em finite} degeneracy, then a downward cusp at
$\nu^*=2$ (and by implication at $\nu=2/5$) would result. Thus,
for general FQHE states, either boundary conditions with strong
confinement or compact geometries are required.

In the end, we note that the edge-wave description of
Ref. \cite {Stone} is applicable  only to the region  close to
$\nu=1$ where there are no cusps (this region grows with $N$).
Lee and Wen have found that the (approximate) degeneracy of the
low-energy states at $3M+L^*$ is equal to that of non-interacting
electrons at $M+L^*$ \cite {Lee};
this is also straightforwardly understandable in the CF approach.

In conclusion, this work demonstrates
that the composite fermion approach provides useful insight into the
results of the exact diagonalization study of quantum dots in high
magnetic fields.  A large number of features are explained
in a simple, economical, and systematic manner.
This work was supported in part by the Office of Naval Research under
Grant no. N00014-93-1-0880, and by the NSF under Grant no. DMR93-18739.

{\bf Figure Captions}

Fig. 1. This figure shows the exact interaction energy of the
lowest energy state as a function of the total angular momentum
$L$ for $N=3$ interacting electrons (solid lines).
The dashed line marked by $m=0$ shows the kinetic energy $K(L^*)$ of
non-interacting electrons with $L^*=L$.
The filled circles on the dashed line indicate
the downward-cusp states.
The dashed curve has been vertically shifted for clarity.

Fig. 2. Same as in Fig. 1 for $N=4$.

Fig. 3. Same as in Fig. 1 for $N=5$ and 6.

Fig. 4. This figure shows the cusp sizes (defined in the text)
measured in units of the size of the first upward cusp (at $L=M+1$)
of the $m=1$ composite fermion \cite {cusp}.  The filled symbols refer to
the cusps of the $m=2$ composite fermions.


\begin{thebibliography}{99}
\bibitem{F5} Not all downward cusp states become ground states. For
example, the states at  $L=$27 and 33 for six electrons, which
correspond to $L^*=$ -3 and 3, do not become
ground states \cite {Yang}. In this case, the CF theory provides
an interesting explanation. Consider $L^*=$ -3. The energy of
this state (in the dashed curve) lies
above the straight lines joining the two downward cusps on either
side, i.e., it would never become the ground state. Same is true for
$L^*=$ 3. This explanation, however, does not always work.
\bibitem{cusp} The size of the upward cusp at $L=M+1$
($\hbar\omega^{CF}$) is given by  0.1486,  0.1348,
0.1240,  0.1153,  0.1081, 0.1020 (in units of $e^2/\epsilon a$) for
$N=$3 to 8, respectively.
\end{thebibliography}
\end{document}